\documentstyle[graphics,epsf]{lamuphys}
\makeatletter
\let\chapter\hid@chapter
\makeatother

\def \etal {{\em et al.}}
\def \eg {{\em e.g.}}
\def \ie {{\em i.e.}}

\def\eps@scaling{.95}
\def\epsscale#1{\gdef\eps@scaling{#1}}

\def\plotone#1{\centering \leavevmode
    \epsfxsize=\eps@scaling\columnwidth \epsfbox{#1}}

\def\plottwo#1#2{\centering \leavevmode
    \epsfxsize=.45\columnwidth \epsfbox{#1} \hfil
    \epsfxsize=.45\columnwidth \epsfbox{#2}}

\begin{document}
\pagenumbering{arabic}

\title{Stellar Populations and Galaxy Morphology at High Redshift}
\titlerunning{Stellar Populations and Galaxy Morphology at High-$z$}
\author{Andrew Bunker\inst{1,2}, Hyron Spinrad\inst{1}, Daniel
Stern\inst{1,3}, Rodger Thompson\inst{4}, Leonidas Moustakas\inst{1,5}, 
Marc Davis\inst{1} and Arjun Dey\inst{6,7}}
\institute{Department of
Astronomy, University of California at Berkeley,\\ 601 Campbell Hall,
Berkeley CA~94720, USA
\and
Institute of Astronomy, Madingley Road, Cambridge CB3~0HA, UK\\
{\tt email: bunker@ast.cam.ac.uk}
\and
Jet Propulsion Laboratory, California Institute of Technology, MS
169-327, Pasadena CA~91109, USA
\and
Steward Observatory, University of Arizona, Tucson AZ~85721, USA
\and
Astrophysics Department, 1 Keble Road, Oxford OX1~3RH, UK
\and
Kitt Peak National Observatory, 950 N.~Cherry Ave., Tucson AZ~85726, USA
\and
Department of Physics \& Astronomy, The John Hopkins University, 
Baltimore MD~21218, USA
}
\authorrunning{Bunker \etal }

\maketitle

\begin{abstract}
In this article we investigate the morphology and stellar populations of
high-redshift galaxies through multi-waveband HST imaging and
ground-based spatially-resolved spectroscopy.  We study the redshift
evolution of galaxy morphology in the Hubble Deep Field, using the deep
IDT-NICMOS near-infrared HST imaging coupled with spectroscopic and
photometric redshifts. Using the multi-waveband data to compare the
appearance of galaxies at the same rest-frame wavelengths reveals that
morphological $k$-corrections (the change in appearance when viewing
high-$z$ objects at shorter rest-frame wavelengths) are only important
in a minority of cases, and that galaxies were intrinsically more
peculiar at high redshift. One example of significant morphological
$k$-corrections is spiral galaxies, which often show more pronounced
barred structure in the near-infrared than in the optical. Therefore,
the apparent decline in the fraction of barred spirals at faint
magnitudes in the optical HDF may be due to band-shifting effects at the
higher redshifts, rather than intrinsic evolution.

Using such features as the age-sensitive Balmer$+$4000\,\AA\ break, the
spatially-resolved colours of distant galaxies in optical/near-infrared
imaging can also be used to study their component stellar
populations. We supplement this with deep Keck/LRIS spectroscopy of two
extended sources: a chain galaxy at $z=2.8$ (HDF\,4-555.1, the ``Hot
Dog'' -- the brightest $U$-drop Lyman-break galaxy in the HDF) and a
pair of $z=4.04$ gravitationally lensed arcs behind the cluster
Abell~2390. The absence of measurable rotation across the $z=2.8$ chain
galaxy implies that it is unikely to be a disk viewed edge on. With the
resolution enhancement from lensing, we detect stellar populations of
different ages in the $z=4$ arcs. The Ly-$\alpha$ emission powered by
the H{\scriptsize II} regions is spatially offset from the star-forming
knots in these arcs, possibly as a result of resonant scattering by
neutral hydrogen.
\end{abstract}

%\keywords{galaxies: evolution --- galaxies: fundamental parameters
%(classification) --- galaxies: interactions --- galaxies: irregular
%--- galaxies: peculiar --- infrared: galaxies}

\section{Introduction}

Until recently, the study of the most distant galaxies was very
restricted: unresolved ground-based imaging gave measurements of the
global colours; noisy spectra revealed AGN- or starburst-driven emission
lines; and WFPC\,2 imaging with the Hubble Space Telescope (HST) showed
the shape of high-redshift galaxies in their unfamiliar rest-frame
ultraviolet light. However, our knowledge has radically increased
through the light-gathering power of the Keck telescopes, and the
availability of the near-infrared NICMOS camera on HST. Deep Keck
spectra showing the rest-ultraviolet continuum and stellar- and
ISM-absorption features in $z\sim 3$ galaxies (\eg, Steidel \etal\
1996b; Pettini \etal\ 2000) has opened the door to spectroscopic study
of stellar populations in the distant Universe. A complementary approach
to exploring the composition of high-redshift objects comes from
multi-waveband HST imaging from the ultraviolet to the near-infrared,
which
%crucially 
provides the resolution necessary to study their {\em
spatially-resolved} stellar populations.

Morphology offers a window on the evolutionary status of
galaxies. However, because of band-shifting effects, the interpretation
of the shape of galaxies hinges on knowledge of their redshifts and
spectral energy distributions (SEDs). In this article we explore this
through high resolution optical/near-infrared imaging and deep
spectroscopy of distant galaxies.  To resolve the issue of whether the
peculiar galaxies which dominate the number counts at faint magnitudes
are the counterparts of local irregulars or whether their morphological
peculiarity is due mainly to band-shifting effects at high redshift, an
unbiased study of the rest-frame optical morphological properties is
demanded, where the effects of dust and recent star formation are less
dominant than in the rest ultraviolet. In
Section~\ref{sec:HDFmorphology} we address this by analysing the
IDT-NICMOS images of the northern Hubble Deep Field (HDF). In
Section~\ref{sec:hotdog} we present a case study of the brightest $z\sim
3$ galaxy in the HDF. This example, HDF\,4-555.1 (known as the ``Hot
Dog''), is a highly elongated system and is sufficiently extended to
allow resolved ground-based spectroscopy, which we have obtained with
Keck/LRIS (Oke \etal\ 1995). We use a similar approach to explore the
stellar populations in a pair of $z\approx 4$ gravitationally-lensed
arcs behind the cluster Abell~2390, which is described in
Section~\ref{sec:abell2390}. Throughout, we assume a cosmology where
$h_{50}=H_{0}~/~50~{\rm km~s}^{-1}~{\rm Mpc}^{-1}$, $q_{0}=0.5$ and
$\Lambda=0$, unless otherwise stated.
{All magnitudes in this paper are with respect to the
AB system (Oke \& Gunn 1983) where $m_{AB} = -48.57 -
2.5\log_{10}~f_{\nu}/{\rm (erg~cm^{-2}~s^{-1}~Hz^{-1})}$.} 

\section{Galaxy Morphology and its Redshift Evolution}
\label{sec:HDFmorphology}

With ground-based seeing, the study of galaxy morphology was restricted
to redshifts of no more than a few tenths.  The advent of HST, and its
resolution of $\sim$~0.1\arcsec, has revolutionized this field. Results
from projects such as the HST Medium Deep Survey (MDS, Griffiths \etal\
1994ab) have shown that at faint magnitudes ($I_{AB}>21$) an increasing
fraction of galaxies do not conform to the traditional categories (\eg,
Glazebrook \etal\ 1995; Driver \etal\ 1995).  The first Hubble Deep
Field (Williams \etal\ 1996) dramatically pushed this study to even
lower fluxes, tracing sub-$L^{*}$ galaxies to high redshift. The optical
images of the HDF show that by $I_{AB}\ga 24$, the conventional Hubble
sequence no longer provides an adequate description of many or most
galactic systems (Abraham \etal\ 1996; Driver \etal\ 1998). Indeed, at
higher redshifts we may be seeing new classes of galaxy emerge with no
local counterpart, such as the `chain galaxies'
(Section~\ref{sec:hotdog} and Cowie, Hu \& Songaila 1995) and `tadpoles'
(van den Bergh \etal\ 1996).
 
Some of these faint sources are intrinsically under-luminous peculiar
galaxies at modest redshift. However, the median redshift has risen to
$z\ga 1$ for a limiting magnitude of $I_{AB}=26$ (Lanzetta,
Fern\'{a}ndez-Soto \& Yahil 1997). Hence, in the faint magnitude
r\'{e}gime, band-shifting effects become important: the optical
passbands
sample shorter rest-frame wavelengths in galaxies at the higher
redshifts, and large ``morphological $k$-corrections'' can arise (\eg,
Odehahn \etal\ 1996). At $z\ga 1$, the appearance in the observed
optical is dominated by regions of recent star formation, luminous in
the rest-frame ultraviolet on account of the massive, short-lived OB
stars.
Indeed, Colley \etal\ (1996) suggest that the observed peak in the
two-point angular correlation function of the optical HDF at $\approx
0\farcs3$ is due to mis-classifying multiple compact star-forming
regions within larger high-redshift galaxies as separate systems,
exacerbated by the cosmological $(1+z)^{-4}$ bolometric
surface-brightness dimming which boosts the contrast between the compact
star-forming knots and the more diffuse host galaxy.

\subsection{High-Resolution Imaging in the Near-Infrared}
 
The rest-optical is a far better tracer of the dynamical mass of a
galaxy than the ultraviolet. This suggests a strategy of high-resolution
imaging in the near-infrared; the $V$- and $R$-bands in the rest-frame
of a $z\approx 1$ galaxy are well approximated by the $J$- and
$H$-passbands, and multi-colour imaging out to the $H$-band can trace
the rest-frame $B$-band morphology of galaxies as far as $z\approx
3$. However, until recently there has been no high-resolution infrared
data set which reaches a limiting flux comparable to the optical HDF.
 
The Instrument Development Team (IDT) of the HST NICMOS camera (Thompson
\etal\ 1998) have imaged an area of the northern HDF to unprecedented
depth in the near-infrared, observing for 49\,orbits in each of the
F110W and F160W filters (centered at $1.1\,\mu$m and $1.6\,\mu$m and
similar to the ground-based $J$- and $H$-bands). The widest-field NIC\,3
camera was used to survey a $\sim 1\, {\rm arcmin}^{2}$ portion of the
HDF.  A detailed description of the observations and data reduction are
given by Thompson \etal\ (1999). Once we correct for different
resolutions of NIC\,3 and WFPC\,2 (through ``PSF matching''), we can use
the spatially-resolved colours to study different stellar populations
and/or dust-reddening within a galaxy (see
Figs.~\ref{fig:spiralSEDs}\,\&\,\ref{fig:spat_res_hotdog}).

\subsection{The Transformation of Spiral Galaxies with Wavelength}

\begin{figure}[ht]
\plottwo{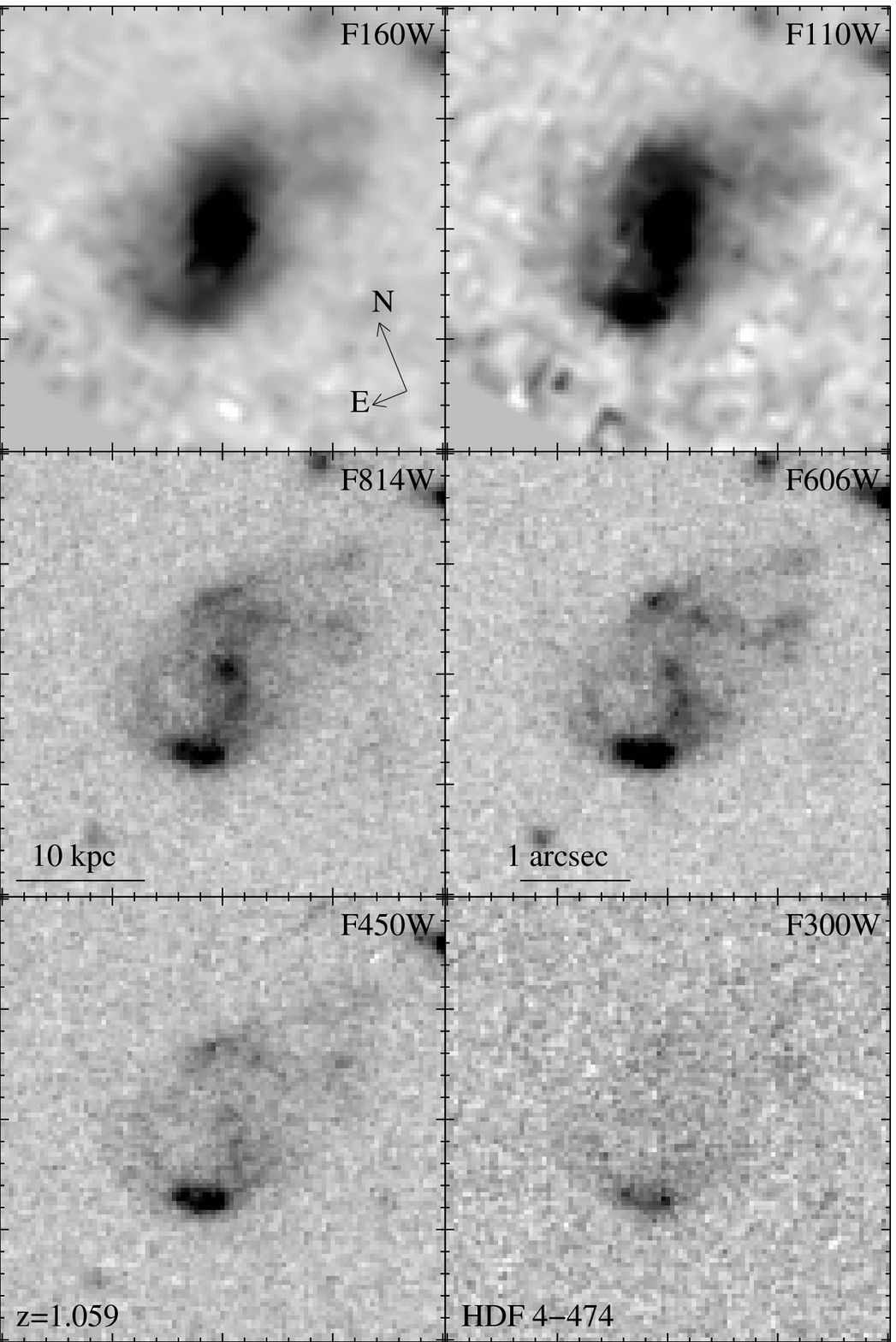}{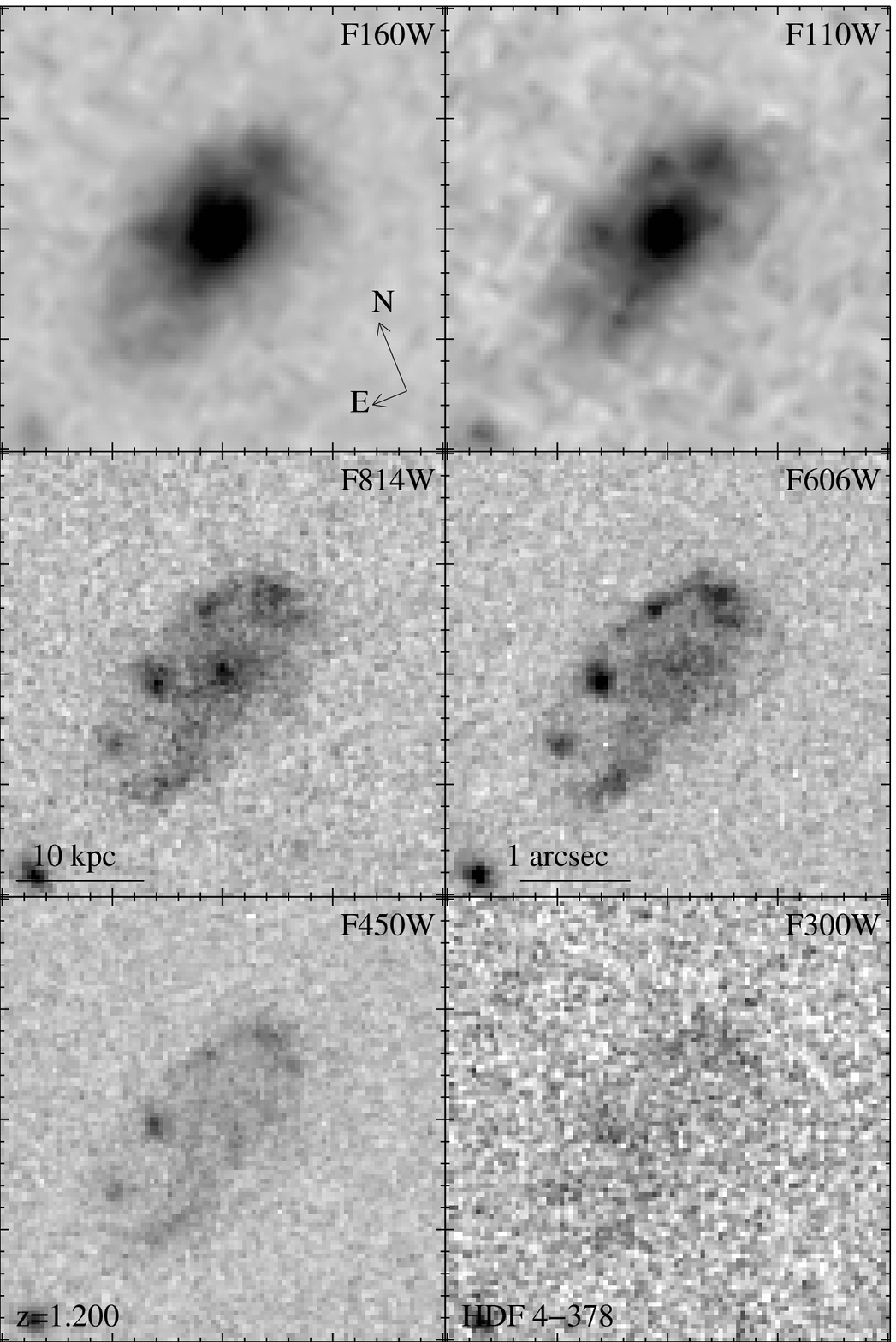}
\put(-290,210){\large \bf a}
\put(-130,210){\large \bf b}
\caption[]{Spiral galaxies at $z\approx 1$, showing the great change in
apparent morphology going from the optical (the rest-ultraviolet, where the
appearance is irregular) to the near-infrared, where their true spiral nature
is revealed. In the case of HDF\,4-474 (left), the WFPC\,2 images are
dominated by a star forming knot, and for HDF\,4-378 (right) the
older/redder population of the bulge is only visible at infrared wavelengths.}
\label{fig:spirals}
%\end{figure}
%\begin{figure}[h]
\plottwo{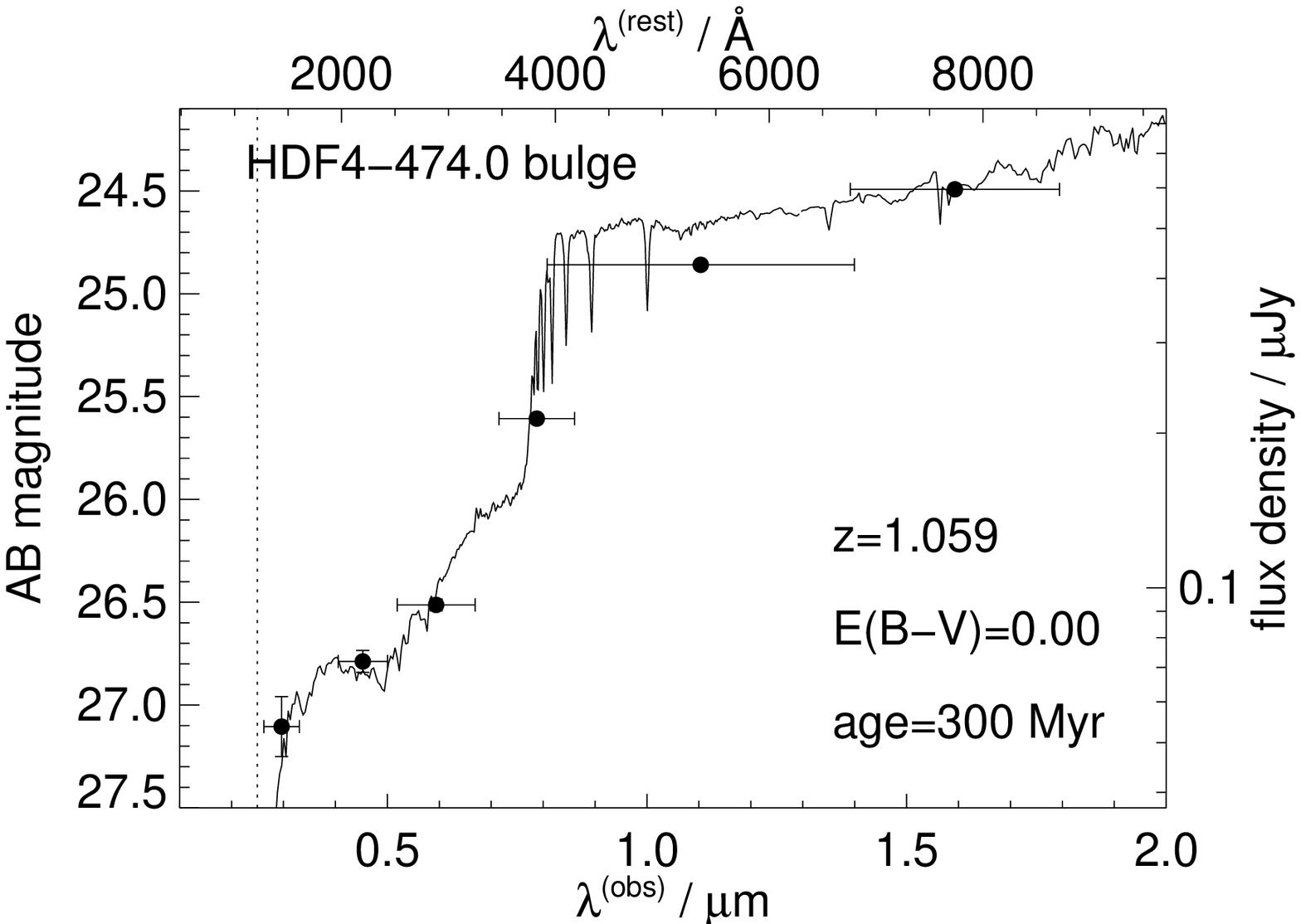}{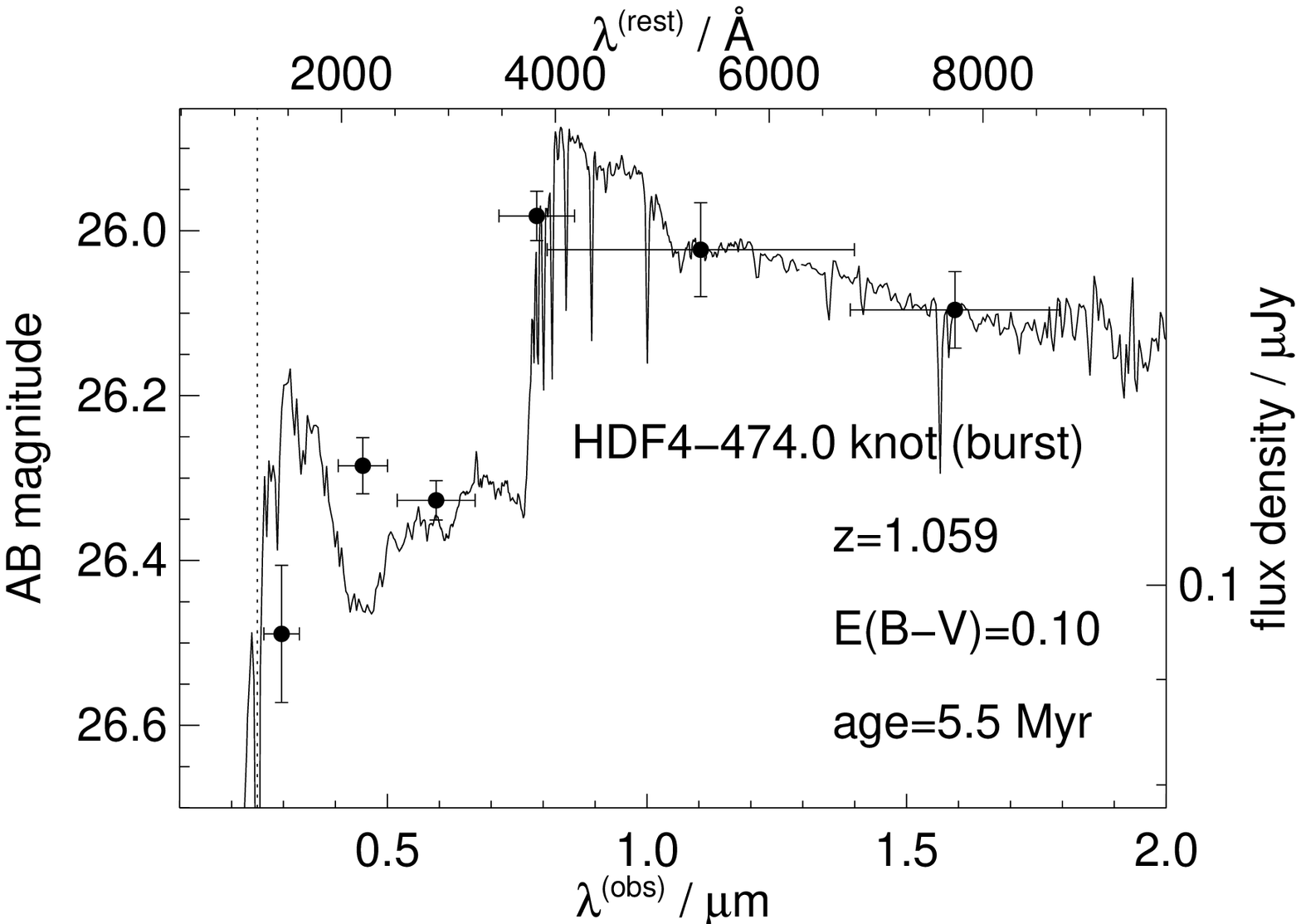}
\caption[]{Stellar population fits to two spatially resolved regions of the
$z\approx 1$ spiral HDF\,4-474 (see Fig.~\ref{fig:spirals}a), using the
latest version of the Bruzual \& Charlot (1993) models. The bulge (left
panel) is clearly very much older than the star-forming H{\scriptsize
II} region in one of the spiral arms (right panel).}
\label{fig:spiralSEDs}
\end{figure}

One of the most visually striking differences between the optical and
near-infrared HDF images are spiral galaxies at moderately-high redshift
($z\sim 1$). At NICMOS wavelengths (the rest-optical), many of these are
clearly classic spirals, and therefore dynamically-evolved stable
systems which certainly should not fall under the banner of
morphological peculiars. However, as illustrated in
Fig.~\ref{fig:spirals}, moving to the rest-UV shifts the classification
toward a much later Hubble type -- \ie, becoming more irregular (Bunker,
Spinrad \& Thompson 1999). In extreme cases, the galaxy appearance is
such a strong function of wavelength that some systems which resemble
small groups of tidally-interacting sub-galactic clumps in the WFPC\,2
optical images are only unveiled as nucleated spirals by the infrared
observations. A classic example is the galaxy HDF\,4-474.0 at $z=1.059$
(Cohen \etal\ 1996) which is totally dominated by an off-centre star
forming H{\scriptsize~II} region in the $U$- and $B$-images, but
transforms into a `grand design' face-on spiral in the near-infrared
(Fig.~\ref{fig:spirals}a). Spiral bulges are dominated by cool giants,
and so brighten at the redder wavelengths; in the case of HDF\,4-378 (at
an estimated photometric redshift of $z=1.20$, Fern\'{a}ndez-Soto,
Lanzetta \& Yahil 1999) the bulge is {\em totally absent} from the
observed optical passbands, but dominates the infrared light
(Fig.~\ref{fig:spirals}b). This is reminiscent of the far-UV 1500\,\AA\
imaging with UIT of the local spiral, M81, presented in O'Connell
(1997).

\begin{figure}[ht]
\plottwo{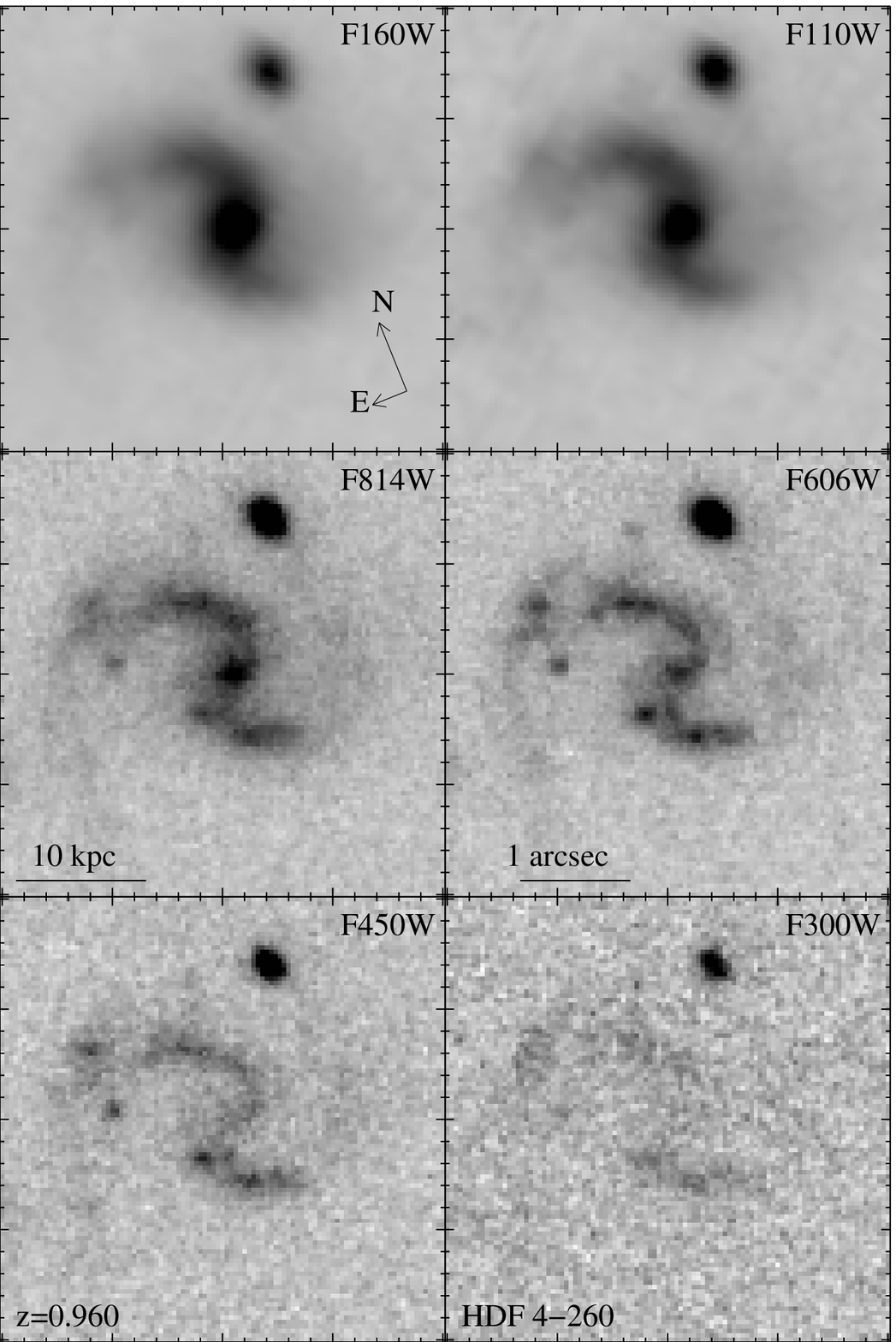}{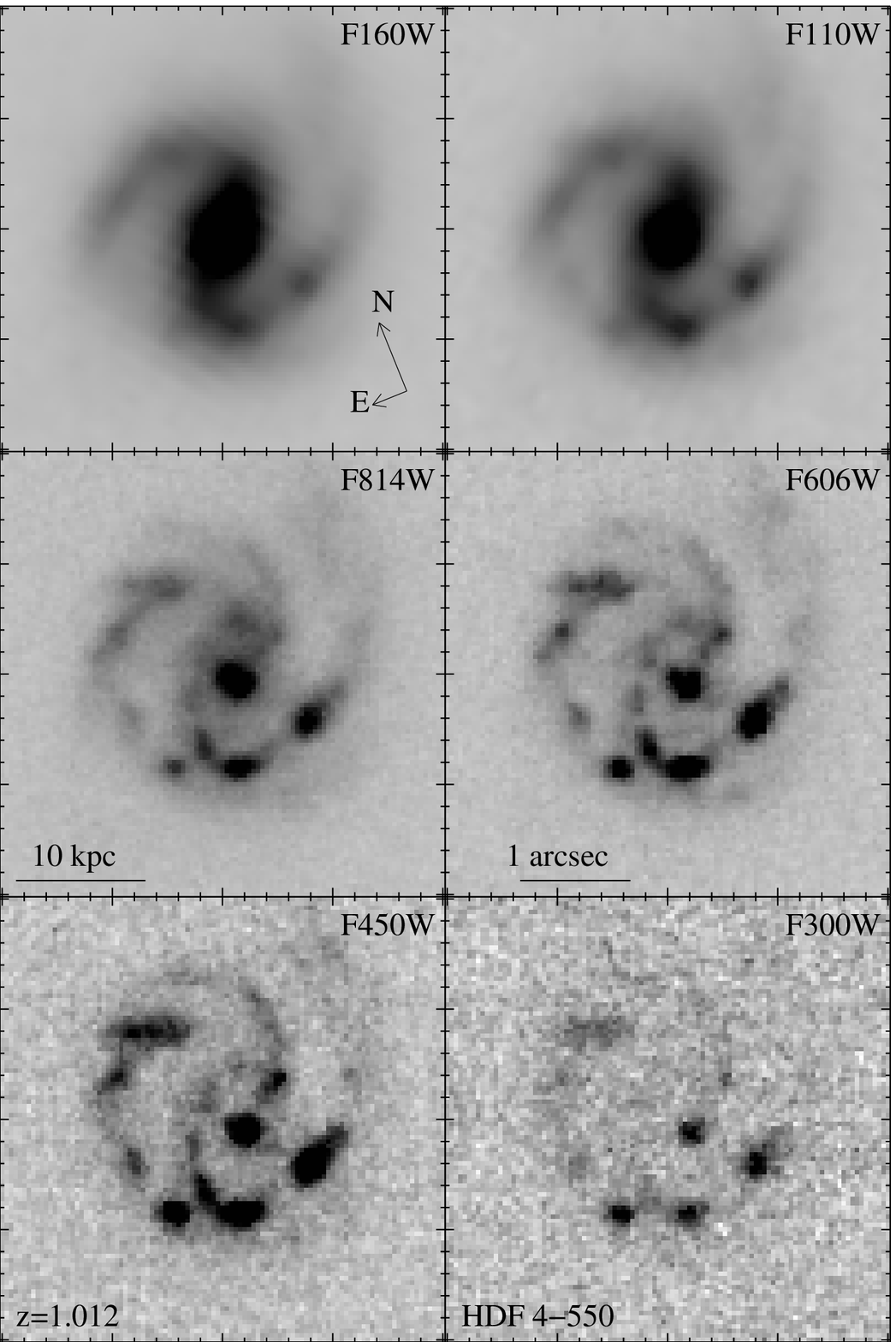}
\put(-290,210){\large \bf a}
\put(-130,210){\large \bf b}
\caption[]{The left panel
shows the only optically-selected barred spiral in HDF-North (van den
Bergh \etal\ 1996), and
this seems to be through chance alignment of a swath of young stars with
the approximate axis of the true bar. The
galactic bar in the spiral displayed in the right panel is only
recognizable at infrared-wavelengths -- at its redshift of $z\approx
1$, the optical wavebands only sample the rest-ultraviolet, where the older \&
redder bulge/bar stellar populations are not prominent.}
\label{fig:barred}
\end{figure}

From the optical HDF, there also appears to be strong redshift evolution
in the relative fraction of galactic bars. Indeed, van den Bergh \etal\
(1996) report just one barred spiral in the whole of HDF-North.  More
recently, Abraham \etal\ (1999) have found similar evolution in the
WFPC\,2 images of HDF-South (Williams \etal\ 1999), with a marked
decline at $z>0.5$ in the proportion of barred spirals in both
fields. If this is a truly evolutionary effect, then it has great
significance for the physics of disk formation.  However, once again the
effects of large morphological $k$-corrections at higher-redshifts makes
the case for evolution inferred from the apparent decline of barred
spirals at faint optical magnitudes less clear cut. Bars are dominated
by older stellar populations, with similar colors to bulges (de
Vaucouleurs 1961), and so are prominent at redder wavelengths. In the
rest ultraviolet, the star forming regions in the disk will typically
dominate the light, and a spiral which would be identified as being
barred when viewed in the rest optical may be (mis-)classified as
unbarred at shorter wavelengths.  Examination of the IDT-NICMOS images
reveals bars in the near-infrared which are undetected in the WFPC\,2
images (\eg, Fig.~\ref{fig:barred}b at $z\approx 1$); hence, claims of
evolution in the frequency of galactic bars based on optical data alone
should be treated with some caution.

\subsection{The Redshift Evolution in the Fraction of Truly Peculiar
Systems}

Using the six wavebands from the WFPC\,2 and IDT-NICMOS imaging of the
Hubble Deep Field, we have compared galaxy morphology at the same
rest-frame wavelengths. Where available, we use the
spectroscopically-measured redshifts (from Cohen \etal\ 1996 unless
otherwise noted). Where no published spectroscopic redshift exists, we
adopt the photometric redshift estimate of Fern\'{a}ndez-Soto, Lanzetta
\& Yahil (1999). Figure~3 of Bunker (1999) shows the rest-frame $B$-band
of all the galaxies in the IDT-NICMOS field brighter than $I_{\rm
AB}=25$, which extends out to $z\approx 3$.

Down to $I_{AB}\approx 25.5$ (the brightest 100 galaxies in IDT-NICMOS
field), only about 1/6 of galaxies change their appearance greatly
between the WFPC\,2 and NICMOS images -- these have large morphological
$k$-corrections. Of the remaining number, about half of the galaxies
retain the same morphology in all wavebands (above the redshifted Lyman
break) and are ``true peculiars''. Hence, the increased fraction of
unusually-shaped systems at faint optical magnitudes is largely due to
evolution rather than simply band-shifting effects.  The remaining third
of galaxies are too compact for changes in morphology to be ascertained
(limited by the NIC\,3 PSF, which has a FWHM of $\approx 0.25$arcsec),
and this fraction increases greatly at magnitudes fainter than $I_{\rm
AB}=25$. For most cosmologies, the higher-redshift systems are on
average more compact, once allowance has been made for the fact that the
higher-redshift systems are intrinsically more luminous in this
apparent-magnitude limited sample.

\section{The ``Hot Dog'' -- A Study of a $z=2.8$ Chain Galaxy in the HDF}
\label{sec:hotdog}

\begin{figure}[ht]
\plottwo{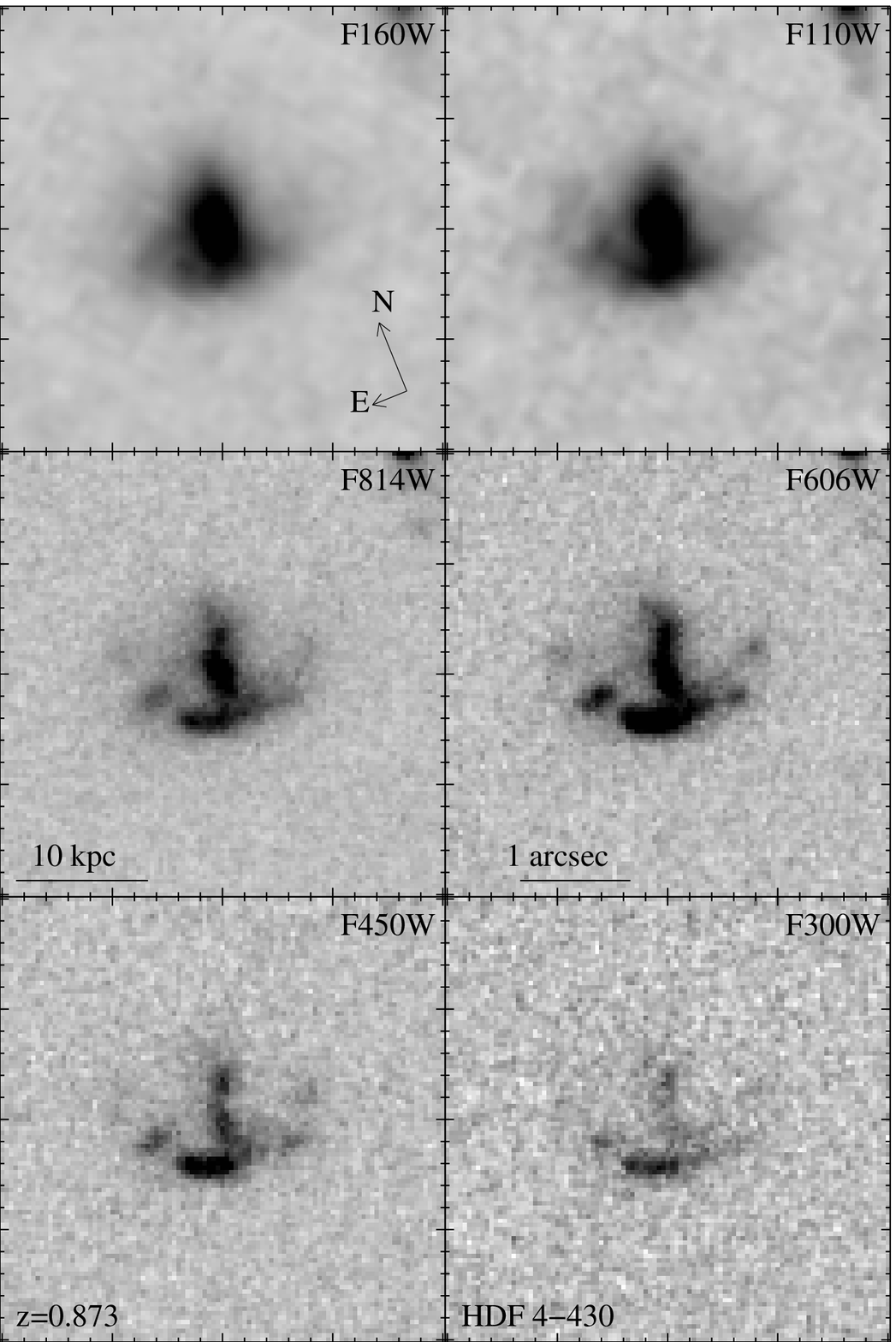}{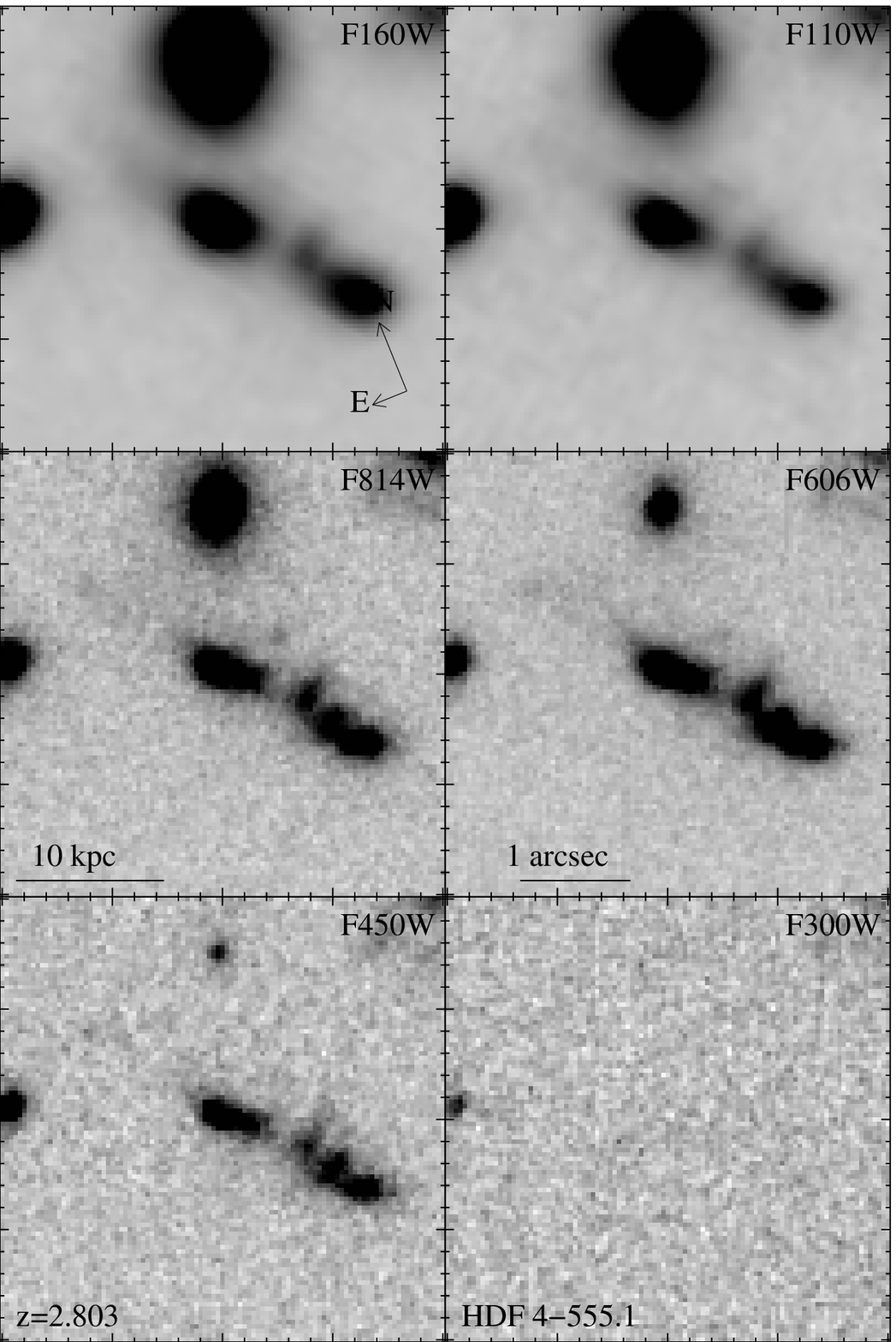}
\put(-290,210){\large \bf a}
\put(-130,210){\large \bf b}
\caption[]{Examples of a bow-shock interacting system (left) and a chain
galaxy (right). Note the bow-shock area itself is comparatively blue,
implying a young stellar population with star formation presumably
triggered by the shock front, whereas the redder (older) core of the
galaxy is more prominent in the near-infrared. The chain galaxy (the
two-component $U$-drop Lyman-break galaxy called ``the Hot Dog'';
Steidel \etal\ 1996a, Bunker \etal\ 1998b) appears the same at all
wavelengths and is blue, implying a relatively homogeneous, young
population (a prim\ae val galaxy candidate?).}
\label{fig:bow_shock_and_chain}
\plottwo{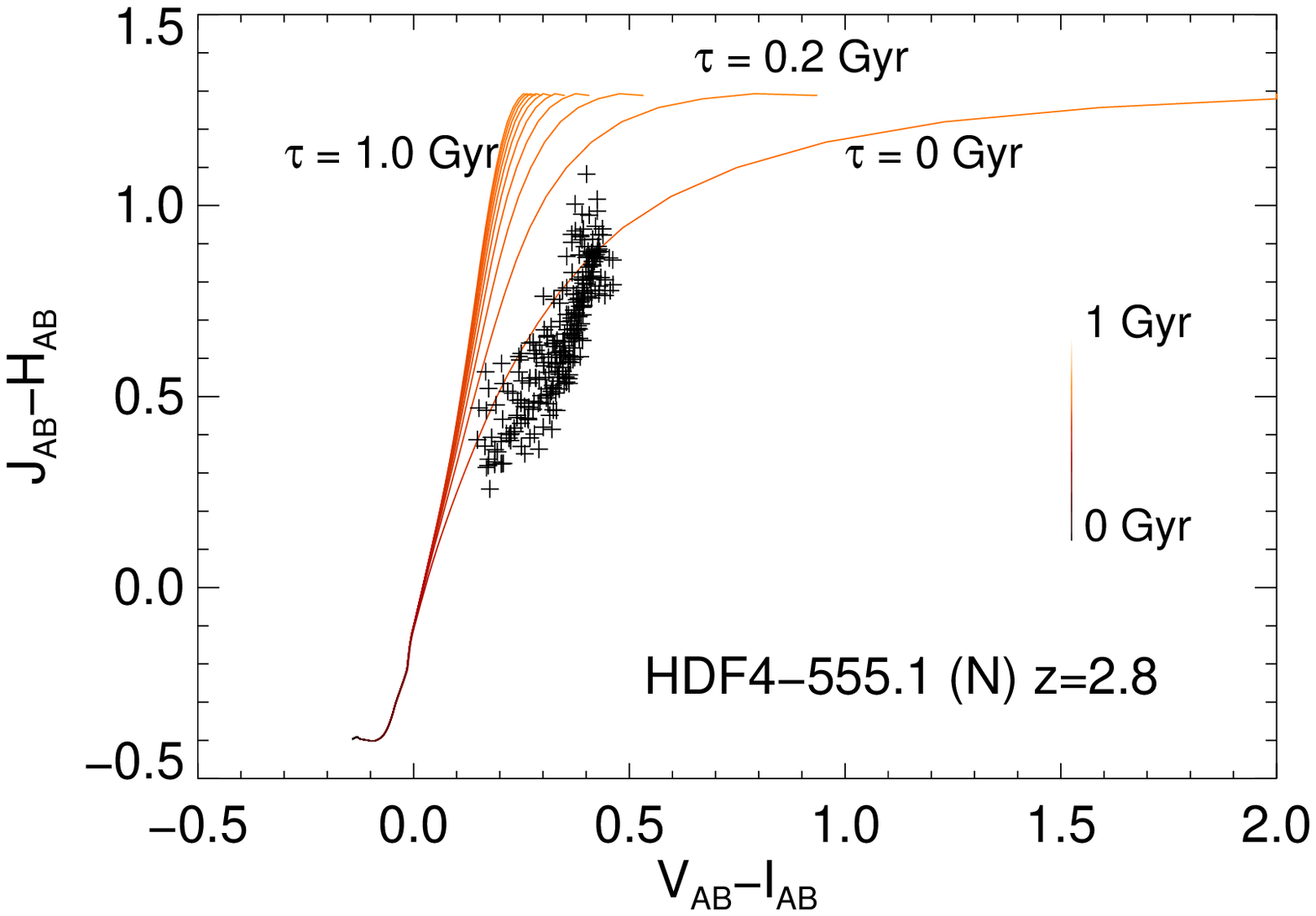}{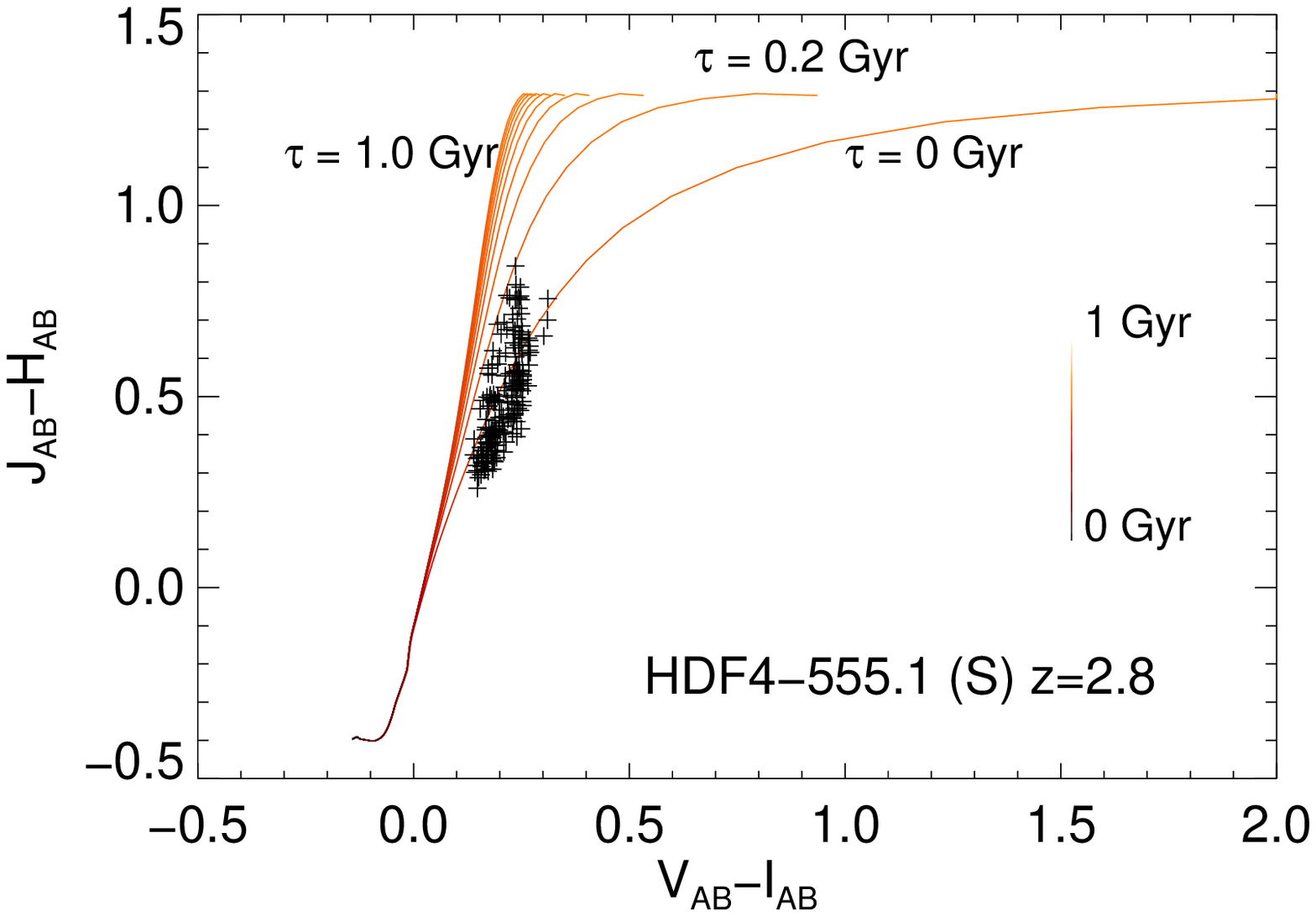}
\caption[]{Spatially-resolved colours of the
northern and southern components of the chain galaxy called ``the Hot
Dog'' (HDF\,4-555.1; Fig.~\ref{fig:bow_shock_and_chain}). The northern and
southern components exhibit subtly 
different colours, attributable to either different stellar populations
or non-uniform dust extinction.  Adopting the approach of Abraham
(1997), we also plot the evolution in the $(V-I)$ and
$(J-H)$ colours with time for a Salpeter IMF and an
exponentially-decaying star formation rate, with $e$-folding times
ranging from 0.1\,Gyr to 1\,Gyr. At $z=2.8$, $(J-H)$ straddles the
age-sensitive 4000\,\AA\ break.}
\label{fig:spat_res_hotdog}
\end{figure}

Some high-redshift galaxies which fall outside the traditional Hubble
tuning-fork diagram belong to new morphological groups, such as tadpoles
(van den Bergh \etal\ 1996) and `bow-shock' systems
(Fig.~\ref{fig:bow_shock_and_chain}). A class which has received much attention
is that of `chain galaxies' (Cowie, Hu \& Songaila 1995). It has been
widely speculated that chain galaxies are linearly-organized giant
star-forming regions, although some have argued that they might be
galactic disks viewed edge-on (\eg, Dalcanton \& Shectman 1996). It is
unlikely that most of the chain galaxies are gravitational lensing
phenomena, as the incidence of potential foreground lenses is small.

The brightest `$U$-drop' galaxy in the HDF is a chain galaxy, quite
unlike the bulk of the Lyman-break population which are
compact and isolated (Giavalisco, Steidel \& Macchetto 1996). This
$z=2.80$ galaxy, HDF\,4-555.1 (see Fig.~\ref{fig:bow_shock_and_chain} and
Steidel \etal\ 1996a, source C4-06), has been dubbed ``the Hot Dog''
because of its highly-elongated morphology -- it is extended over
$\approx 2\farcs5$, and clearly resolved even in ground-based seeing.

Examining the spatially-resolved colors of the Hot Dog reveals that both
of the prominent lobes are well fit with 
young stellar populations ($<100$~Myr), although the southern component is
bluer on average and exhibits a much smaller dispersion in colors than
the northern (Fig.~\ref{fig:spat_res_hotdog}). This indicates that the
star formation history of the northern component is more extended in
time, or that the dust extinction along its length varies more
than for the southern lobe.
There is little evidence for a significant
underlying older stellar population, which might be expected in a disk
viewed edge-on.

We have obtained deep, spatially-resolved optical spectroscopy with
\newline
Keck/LRIS, using a long slit aligned along the major axis of the Hot Dog
(Bunker \etal\ 1998b).  Our 14\,ksec spectra
($\lambda/\Delta\lambda_{\rm FWHM}\sim
1000$) sample the rest-frame ultraviolet ($1090 - 1890$\,\AA ), a region
devoid of strong forbidden emission lines. However, many of the
resonance lines associated with hot stellar winds do appear, although
their P~Cygni profiles are less extended in velocity than those observed
in some local star-bursts. The overall rest-ultraviolet continuum also
indicates that the Hot Dog is an actively star-forming galaxy, albeit
with internal dust extinction of $E(B-V)\approx 0.1^{m}$.
The presence of high-ionization N{\scriptsize~V} and He{\scriptsize~II}
in emission demands at least some O-stars. Ly-$\alpha$ emission is
completely suppressed, and the absorption profile can be fit by a
combination of stellar photospheric absorption and a modest interstellar
hydrogen column of $N({\rm H{\scriptsize~I}})\approx 10^{20}\,{\rm cm}^{-2}$ (a
borderline damped system). The Hot Dog exhibits some of the strongest
interstellar absorption features seen in the Lyman-break
population. However, a search for velocity gradients in these lines
along the major axis of this chain galaxy revealed that any systemic
rotation must be small ($<100\,{\rm km\,s}^{-1}$), inconsistent with an
edge-on rotating disk.

We also serendipitously discovered in our long-slit spectroscopy a
compact companion galaxy, HDF\,4-497.0, with Ly-$\alpha$ emission
($W_{0}\approx 30$\,\AA ) at a redshift within $1000\,{\rm km\,s}^{-1}$
of the Hot Dog, and a projected separation of $35\,h_{50}^{-1}$\,kpc.
The flux in the Ly-$\alpha$ line of this companion galaxy is $3\times
10^{-17}\,{\rm erg\,cm}^{-2}\,{\rm s}^{-1}$, and the rest-ultraviolet
continuum suggests a star formation rate of $SFR_{\rm UV}\approx
5.6\,h_{50}^{-2}\,M_{\odot}\,{\rm yr}^{-1}$. The existence of this
companion closely aligned along the major axis of the chain galaxy
offers support to the contention that this is an example of
star formation triggered by collapse along a filament.

As we have spatially-resolved spectroscopy of an extended high-redshift
source, we can also use our data as a probe of the dimensions of
intervening absorbers without being restricted to the one-dimensional
sightlines avalable from QSOs. Our spatially-resolved spectroscopy
reveals Mg{\scriptsize ~II} $\lambda\lambda$ 2796/2803\,\AA\ absorption
by a foreground system at $z=1.239$. This absorption is most likely
associated with the galaxy HDF\,4-516.0 which has a photometric redshift
consistent with the $z=1.239$ absorption lines, and is at a projected
distance of 2\,arcsec from the Hot Dog.  The Mg{\scriptsize ~II}
absorption is only pronounced in the northern component of the Hot Dog,
and the absence of strong absorption in the southern component enables
us to constrain the physical size of the Lyman limit system (the
optically-thick halo) of this galaxy to be $r<27\,h_{50}^{-1}$\,kpc.

\vspace{-0.3cm}
\section{Resolving the Stellar Populations in a Lensed Galaxy}
\label{sec:abell2390}

Gravitational lensing can be used as a tool to increase the resolution
attainable in studies of distant galaxies. Although morphological
information is hard to disentangle because of the geometric distortions
and uncertainties in lens modelling, the amplification afforded by
strong lensing can allow the stellar populations to be mapped on the
sub-kpc scale, as well as magnifying the total flux.

\subsection{Optical/Near-Infrared Imaging of Lensed Arcs at $z=4.04$}

\begin{figure}[ht]
\plottwo{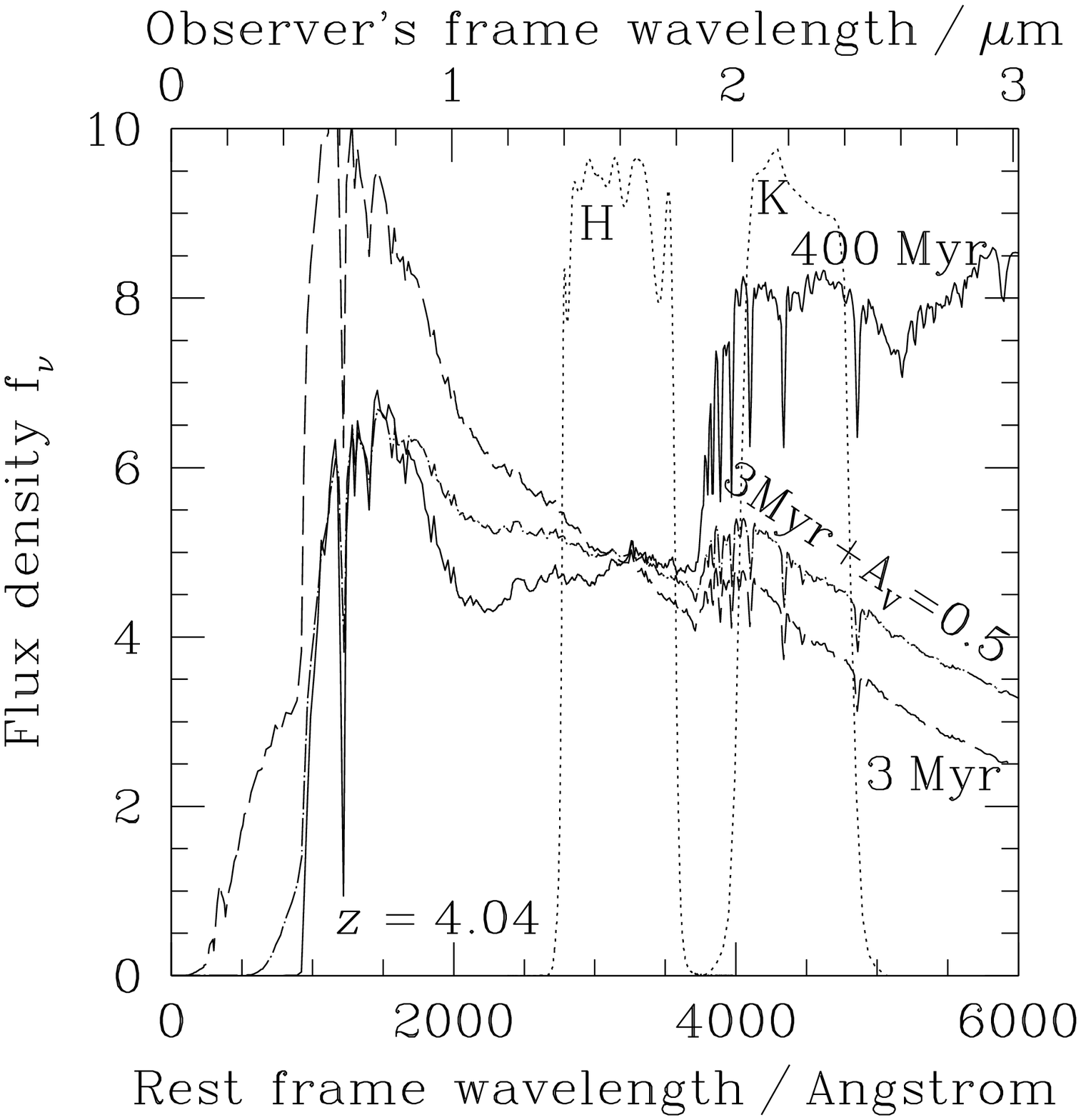}{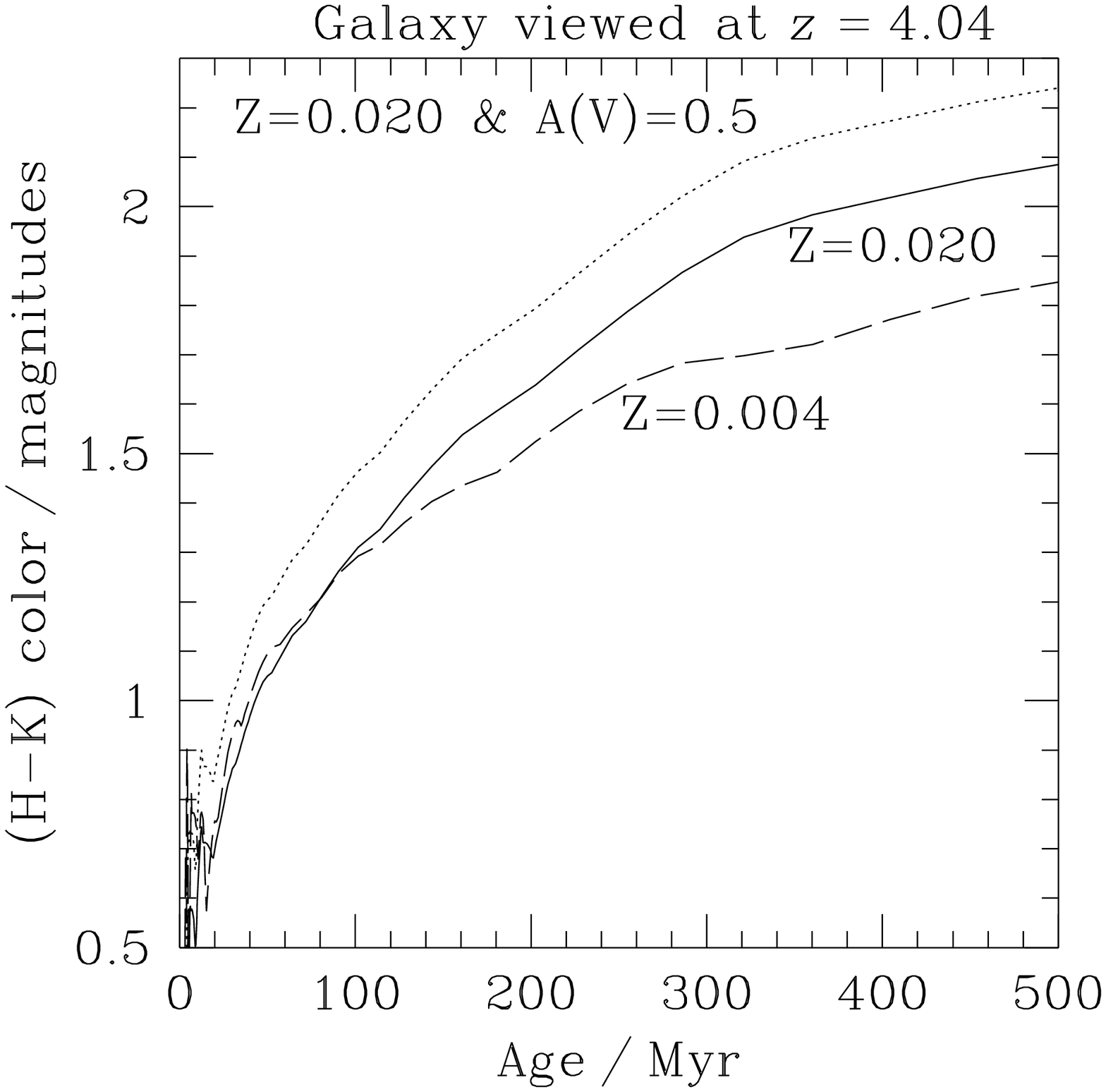}
\caption[]{\underline{Left:} An illustration of
the unreddened rest-frame optical spectra of two galaxies, one observed
only 3\,Myr after the end of an instantaneous burst of star formation
(long-dash curve) and the other seen after 400\,Myr have elapsed (solid
line). We also show the 3\,Myr model with dust extinction of
$A_{V}=0.5^{m}$,
typical of high-$z$ star-forming galaxies (\eg, Pettini \etal\ 1998;
Steidel \etal\ 1999). Note the strong Balmer\,+\,4000\,\AA\ break due to
the older stars. Also plotted (dotted lines) are the $H$ and $K$ filters
in the rest-frame of a $z=4.04$ galaxy, straddling the break. The SEDs
come from the latest Bruzual \& Charlot models.
\label{fig:4000AngBreak}
\underline{Right:} The evolution of the $(H-K)$ colour of a galaxy at
$z=4.04$ as a function of the time elapsed since an instantaneous burst
of star formation. The solid curve is the Bruzual \& Charlot model for
Solar metallicity ($Z=0.020$), with the dashed line showing lower
metallicity, $\frac{1}{5}$ solar ($Z=0.004$).  For this redshift, the
$(H-K)$ colour is an excellent tracer of the time elapsed since the end
of star formation. The dotted curve is the solar-metallicity model with
dust reddening of $A_{V}=0.5^{m}$. \label{fig:4000AngEvln}}
\end{figure}

Combining archival HST/WFPC\,2 data with deep near-infrared imaging
taken with Keck/NIRC (Matthews \& Soifer 1994) in good seeing, we have
measured the spatially-resolved colours in a $z=4.04$ galaxy,
gravitationally lensed by the rich cluster Abell~2390 ($z\approx 0.23$)
into a pair of highly-magnified near-linear arcs 3--5\arcsec\ in length
(Frye \& Broadhurst 1998).  At the redshift of these arcs, the $H$
($\lambda_{\rm cent}\approx1.65\,\mu$m) and $K$ ($\lambda_{\rm
cent}\approx 2.2\,\mu$m) near-infrared pass-bands straddle the
age-sensitive rest-frame 4000\,\AA\,$+$\,Balmer break
(Fig.~\ref{fig:4000AngBreak}).  Comparison of the optical and
near-infrared photometry with a suite of spectral evolutionary models
(the latest version of Bruzual \& Charlot 1993) has enabled us to map
the underlying stellar populations and differential dust extinction
(Bunker \etal\ 1998a). The WFPC2 images clearly reveal several knots,
bright in the rest-ultraviolet, which correspond to sites of active star
formation. However, there are considerable portions of the arcs are
significantly redder, consistent with being observed $> 100$Myr after
star formation has ceased, with modest dust extinction of $E(B-V)\approx
0.1^{m}$. There is degeneracy in the models between dust reddening and
age for the optical/near-infrared colours, but the most extreme scenario
where the colour gradients are solely due to heavy dust reddening of an
extremely young stellar population are strongly ruled out by upper
limits in the far-infrared/sub-mm from ISO/SCUBA (L\'{e}monon \etal\
1998; Blain \etal\ 1998).

\subsection{Keck/LRIS Spectroscopy}
 
\begin{figure}[ht]
\plotone{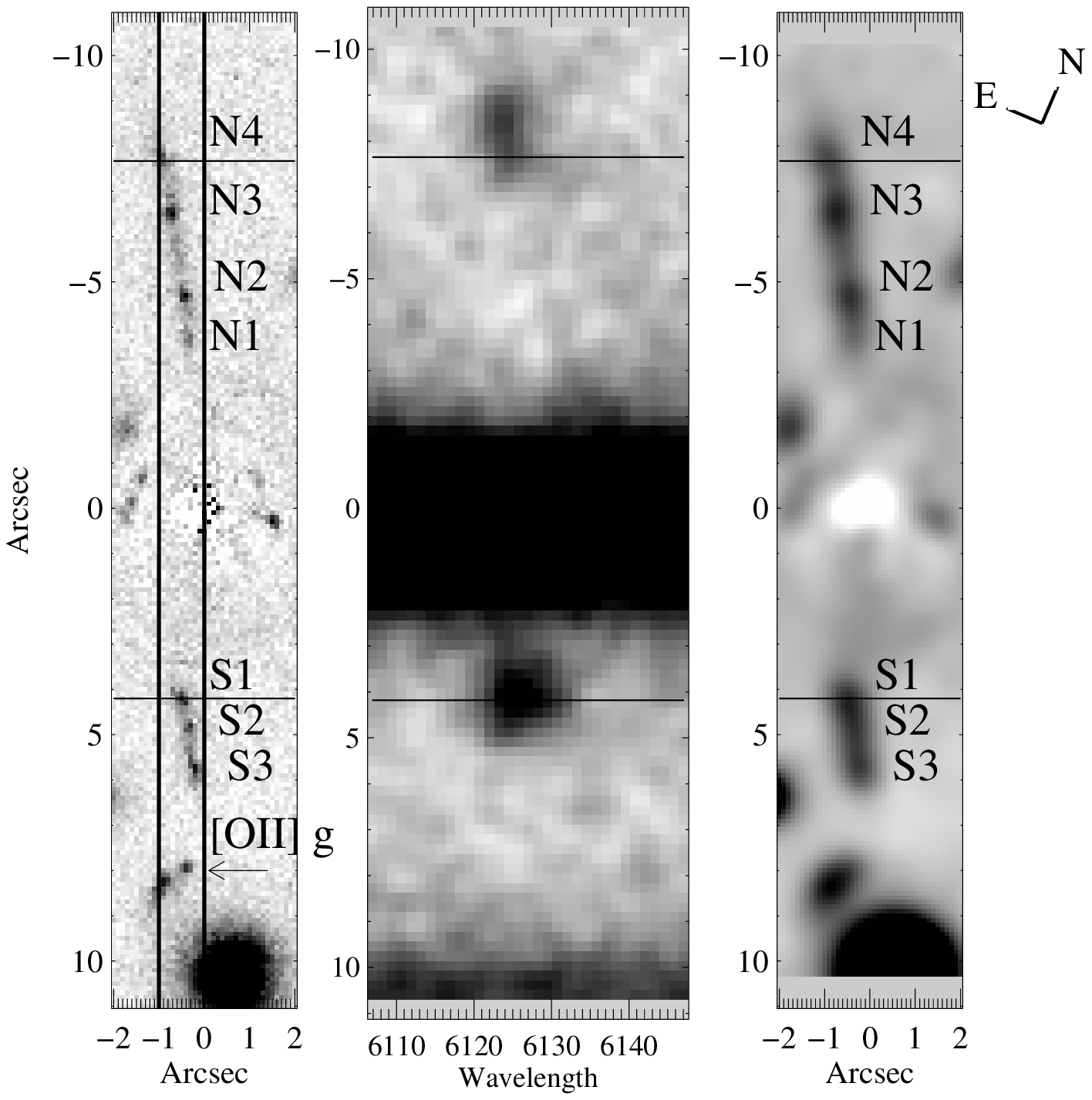}
\caption[]{\underline{Left:} the F814W image with elliptical galaxy
model subtracted (note the counter arcs perpendicular to the axis of the
main arcs, predicted by the lens model of Frye \& Broadhurst 1998). The
area covered by the long-slit optical spectroscopy is shown (slit axis
is vertical). The \underline{right} panel is this elliptical-subtracted
image, smoothed to $\approx 0.6''$ seeing. \underline{Center:} the LRIS
spectrum, with the long-slit aligned along the arcs. The dispersion axis
is horizontal, with wavelength increasing from left to right. The image
has been smoothed by convolving with a Gaussian kernel of
$\sigma=1$\,pixel.  Note the spatial range of Ly-$\alpha$ emission,
which extends well beyond the detectable continuum of the arcs. The
positions of the closest continuum knots N4 ($-7.6''$) and S1 ($4.3''$)
are indicated by the horizontal bars. The knot N4 (top) lies on the edge
of the slit, hence the slight spatial offset of the Ly-$\alpha$ line
centroid and the lower flux compared to the line from S1 (bottom). Also
indicated on the left panel is the $z=1.129$
[O\,II]\,$\lambda$\,3727\,\AA\ galaxy also falling on our spectroscopic
long-slit (see
Fig.~\ref{fig:1DspecLinesAbell2390}). \label{fig:2DspecArcs}}
\end{figure}

We have obtained optical spectroscopy from Keck/LRIS at moderate
dispersion ($\lambda/\Delta\lambda_{\rm FWHM}\approx 1000$) with a long
slit aligned along the major axis of the arcs (Fig.~\ref{fig:2DspecArcs}).  Our
4\,ksec spectrum shows regions with Ly-$\alpha$ in emission that are
adjacent to some of the bright knots seen in the optical HST images
which sample the rest-frame ultraviolet (Figs.\
\ref{fig:2DspecArcs}\,\&\,\ref{fig:1DspecLinesAbell2390}). The non-detections of
N{\scriptsize~V}\,1240\,\AA, C {\scriptsize~IV}\,1549\,\AA\ \&
He{\scriptsize~II}\,1640\,\AA\ strongly favor the interpretation that
the Ly-$\alpha$ arises from the Lyman continuum flux produced by OB
stars, rather than the harder ultraviolet spectrum of an AGN. We see the
Ly-$\alpha$ line morphology extending $\approx 1''$ beyond the
ultraviolet continuum, which we attribute to resonant scattering from
H{\scriptsize~I}
(Bunker, Moustakas \& Davis 2000).  In the bright knots, the SEDs are
consistent with a very young stellar population ($<10$\,Myr) or ongoing
star formation.
 
\begin{figure}[ht]
\plotone{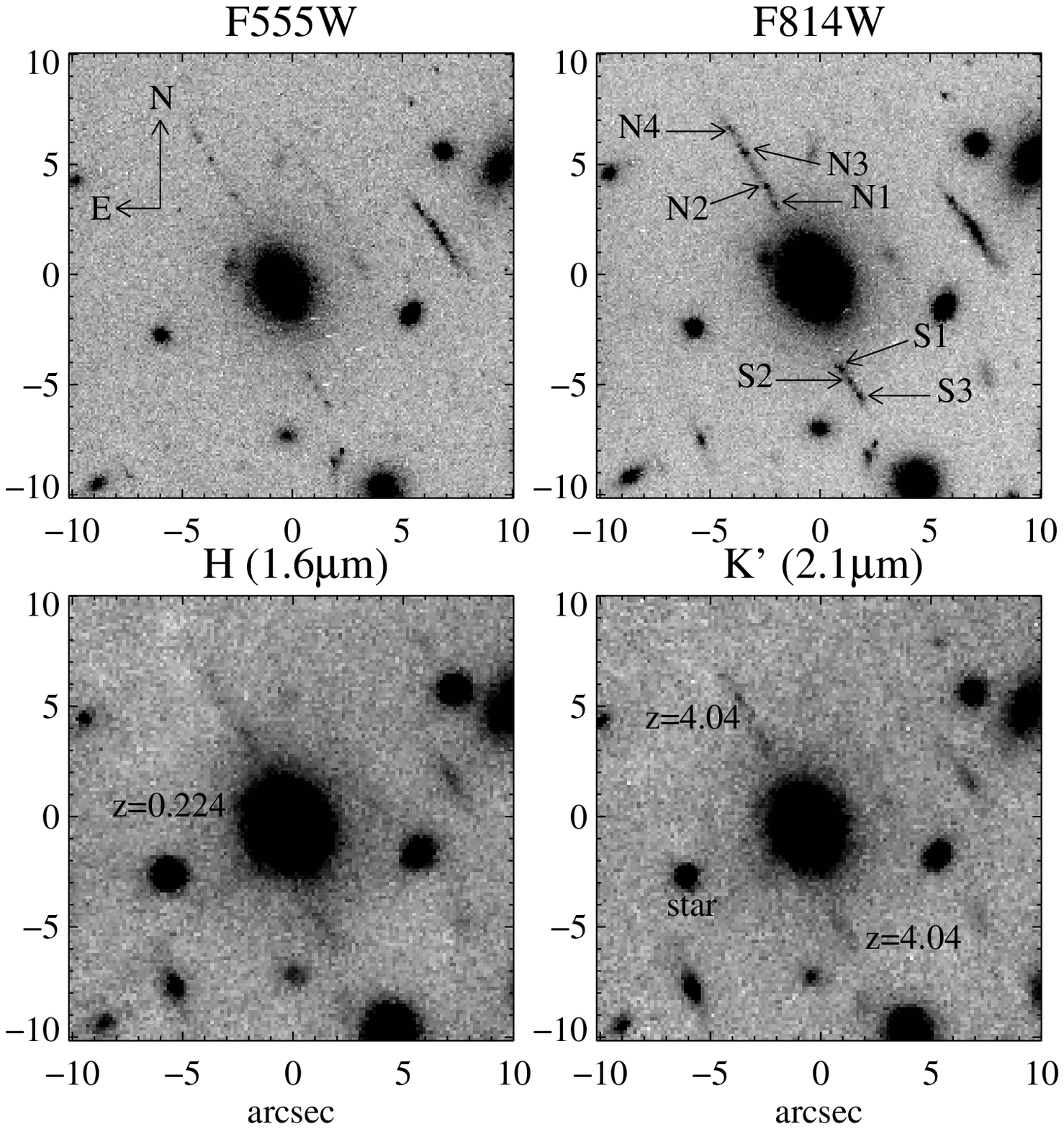}
\caption[]{The top panels show archival HST/WFPC\,2 imaging
   of the cluster Abell 2390.  The $z=4.04$ galaxy is the arclet at
   PA=$+23^{\circ}$ that is bisected by the elliptical.  Top left is the
   HST $V$-band (F555W, 8400\,s) which encompasses Ly-$\alpha$, with the
   HST $I$-band (F814W, 10500\,s) top right. The knots which are bright
   in the rest-ultraviolet (and so are presumably sites of recent star formation)
   are indicated.  Our Keck/NIRC images were obtained in good seeing
   ($0.4-0.5''$ FWHM) and are shown lower left ($H$, 2280\,s) and lower
   right ($K'$, 2880\,s). \label{fig:imageAbell2390}}
\end{figure}
 
\subsection{Evolutionary Status of the $z=4$ Galaxy}

We have evidence for both ongoing star formation and regions of older
stellar populations in the lensed arcs.  It is therefore unlikely that
this $z=4$ system in a true `prim\ae val' galaxy, viewed during its
first major burst of star formation. Rather, our results suggest that
the star formation history of this system has not been coeval, with
current activity concentrated into small pockets within a larger, older
structure.  Correcting for the gravitational amplification (estimated to
be $\approx 10$ from lens models), the intrinsic properties of the
$z=4.04$ galaxy are comparable to the Lyman-break selected $z\approx
3-4$ population of Steidel \etal\ (1996b,1999).  The current
extinction-corrected star formation rate ($\approx
15\,h_{50}^{-2}\,M_{\odot}\,{\rm yr}^{-1}$ for $q_{0}=0.5$) may be
adequate to `build' an $L^{*}$ galaxy over a Hubble time, but a more
likely scenario may be the creation of a sub-unit which will undergo
subsequent merging with nearby systems (such as the other $z=4.04$
galaxy identified in this field by Pell\'{o} \etal\ 1999) to assemble
hierarchically the massive galaxies of today.
 
\begin{figure}[ht]
\plottwo{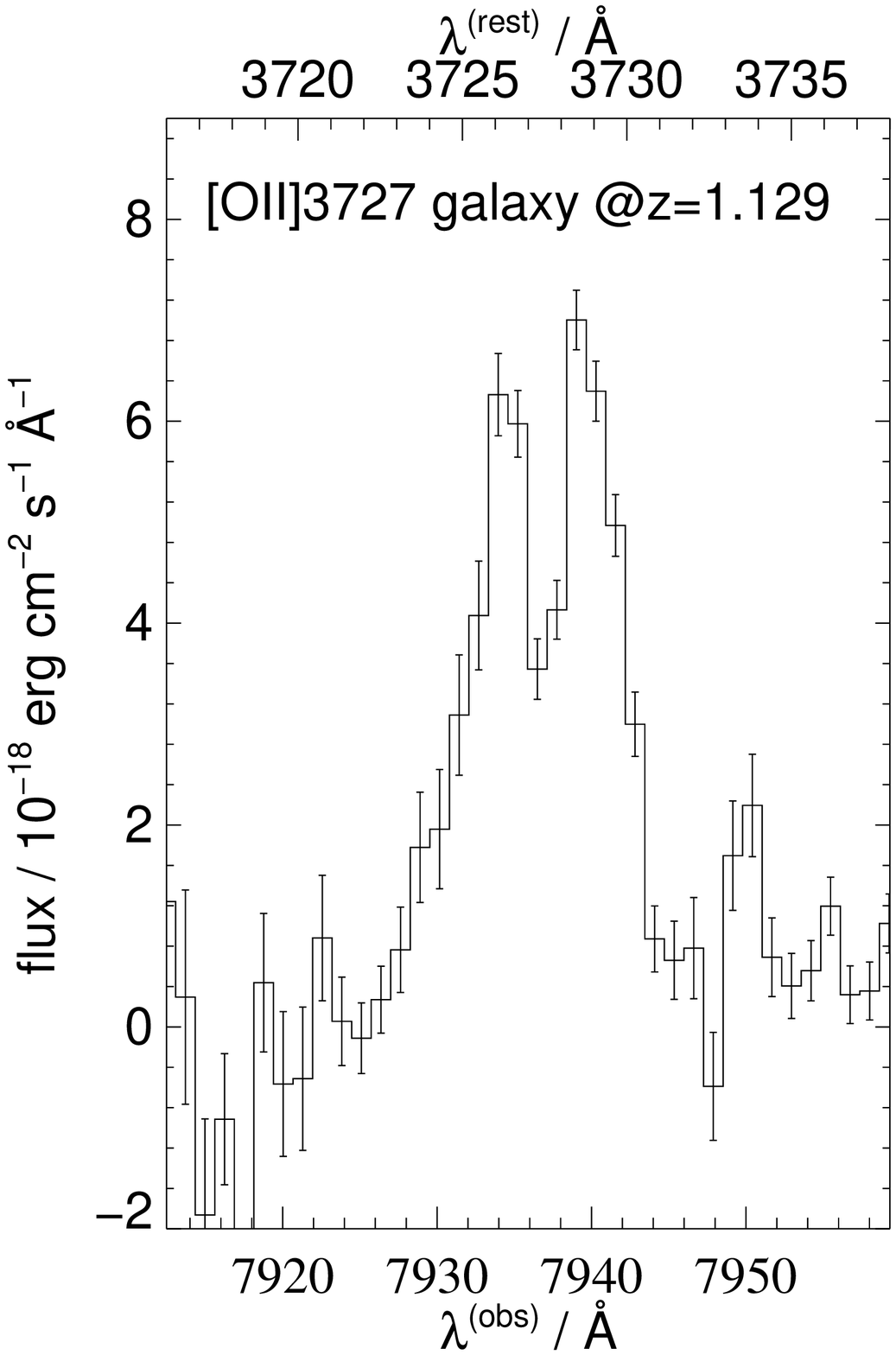}{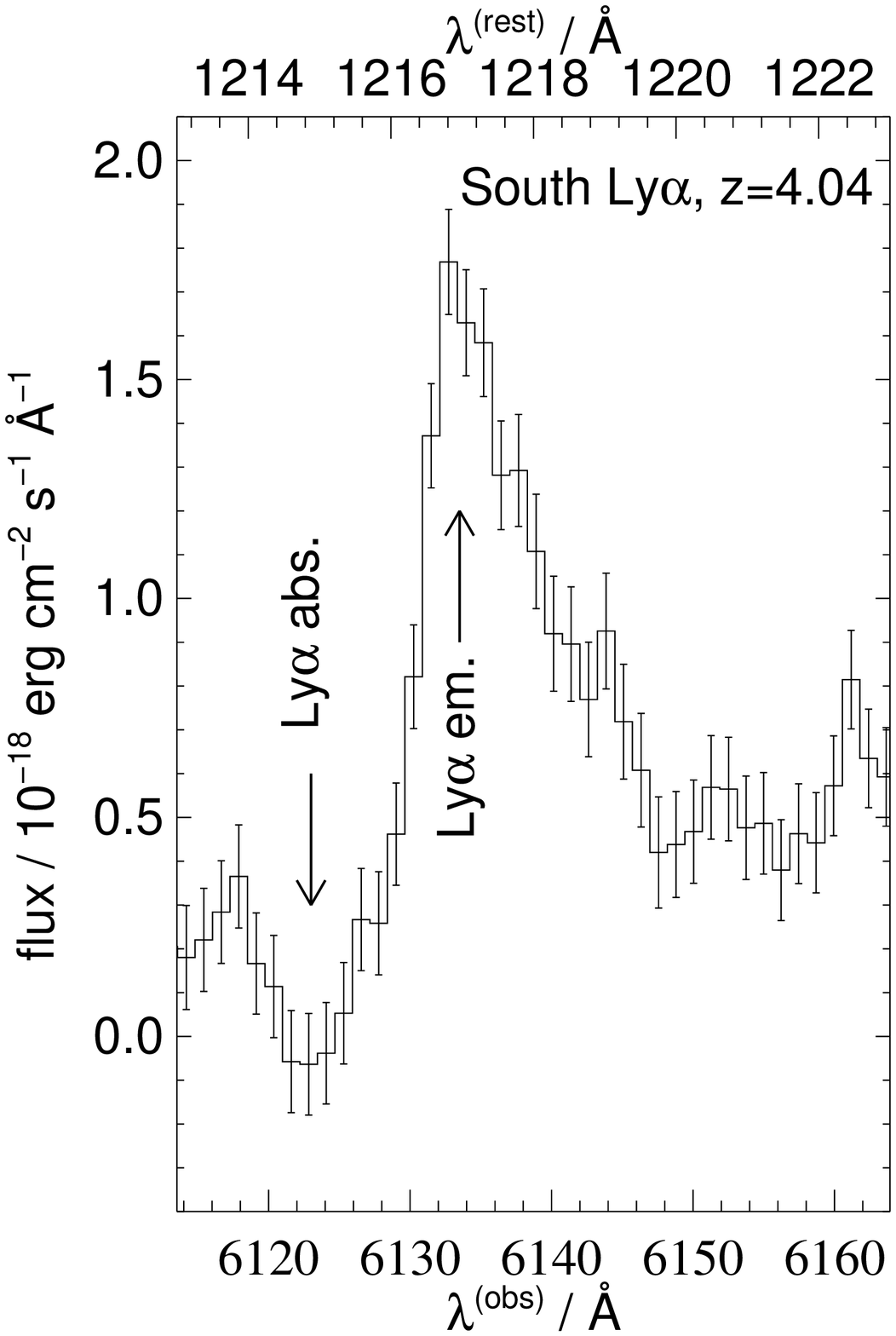}
\caption[]{\underline{Left:} one-dimensional spectral extraction of the
[O{\scriptsize~II}]\,$\lambda\lambda$\,3726.1,3728.9\,\AA\ galaxy at $z=1.1293$,
3$''$ below the southern arc and intercepted by our spectroscopic
long-slit (Fig.~\ref{fig:2DspecArcs}). The extraction width is 7\,pixels
(1.5''). Note that the emission-line doublet is clearly resolved. We do
not see this structure in the emission lines from the arcs, implying
that their origin is not [O{\scriptsize~II}]\,3727\,\AA\ at $z=0.64$.
\underline{Right:} one-dimensional spectral extraction of the southern
arc, showing the region around Ly-$\alpha$ at $z=4.04$. The extraction
width is 8\,pixels ($1.7''$), and encompasses both line- and
continuum-emission regions. The asymmetric emission line profile is
readily apparent, with the sharp decline on the blue side due to
absorption by H{\scriptsize~I} within the galaxy -- a blueshifted
Ly-$\alpha$ absorption trough is visible from the outflowing
H{\scriptsize~I}.
\label{fig:1DspecLinesAbell2390}}
\end{figure}
 
\begin{figure}[ht]
\plotone{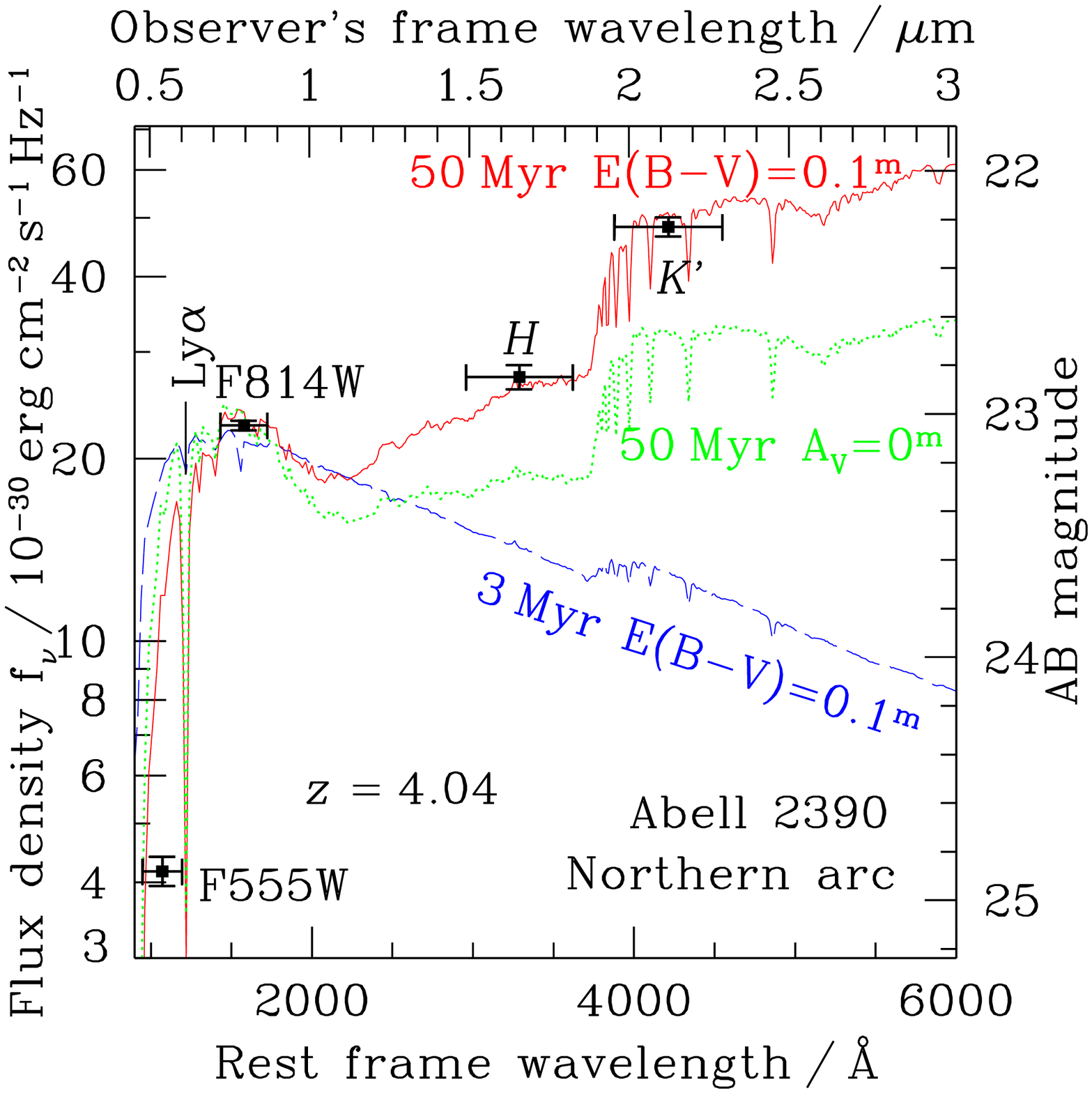}
\caption[]{The broad-band
optical/near-infrared flux from the entire northern arc. Also plotted are
reddened instantaneous-burst stellar population models viewed at various
ages, arbitrarily normalized to the flux measured from the HST/WFPC\,2
F814W image. The flux in F555W is severely attenuated by the opacity of
the intervening Ly-$\alpha$ forest.  The colours are best reproduced by a
stellar population $\sim 50$\,Myr old, with {\em in situ} dust reddening
of $E(B-V)\approx 0.1^{m}$.  Note that at $z=4.04$, the strong
Balmer\,+\,4000\,\AA\ break due to the older stars lies between the $H$-
and $K'$-filters. \label{fig:Abell2390SEDall}}
\end{figure}
 
\section{Conclusions}

In this article, we have explored spatially-resolved stellar populations
at high-redshift, and addressed the impact on studies of galaxy
morphology.  The deep, high-resolution IDT-NICMOS near-infrared imaging
of a portion of the northern Hubble Deep Field has been combined with
the WFPC\,2 data and photometric redshift estimates to study the
redshift evolution of morphology, comparing galaxy appearance at the
same rest-wavelengths.  Some Hubble tuning-fork galaxies only reveal
their true morphology in the near-infrared images. This is particularly
so for galaxies with a large dispersion in stellar ages and
spatially-distinct stellar populations, such as spiral galaxies which
sometimes exhibit galactic bars in the NICMOS images which are invisible
at shorter wavelengths.  However, galaxies which do undergo a
morphological metamorphosis from the WFPC\,2 to NIC\,3 images are in the
minority; most galaxies retain the same appearance in all wavebands, or
are too compact for the structural parameters to be determined. Once the
morphological $k$-corrections have been accounted for, it appears that
the fraction of galaxies falling outside the Hubble sequence does
increase at faint magnitudes/high-$z$.  Many of these ``true peculiars''
show evidence of being dynamically disturbed (possibly through mergers)
with recent star formation activity. From the HST imaging and resolved
spectroscopy with Keck/LRIS, we have shown that a $z=2.8$ chain galaxy
in the HDF has a predominantly young stellar population and no
significant rotation, and is thus unlikely to be an edge-on disk galaxy.
Using gravitational amplification to increase our resolution, we have
also resolved the stellar populations on sub-kpc scales in a system of
$z=4.04$ lensed arcs.

The analysis of galaxy morphologies and colours in multi-waveband
imaging, coupled with resolved spectroscopy, provides a valuable probe
into the stellar populations and evolution of galaxies.  A natural
progression is to use the integral field unit spectrographs currently
being developed. Spectroscopy of spatially-resolved stellar populations
in the high-redshift Universe will be a major scientific goal for NGST
and the next generation of large ground-based telescopes with adaptive
optics.

\section*{Acknowledgments}

 AJB acknowledges a NICMOS postdoctoral fellowship while at Berkeley
 (grant NAG\,5-3043), and a U.K.\ PPARC observational rolling grant at
 the Institute of Astronomy in Cambridge
 (ref.~no.~PPA/G/O/1997/00793). The observations were obtained in part
 with the NASA/ESA Hubble Space Telescope operated by the Space
 Telescope Science Institute manged by the Association of Universities
 for Research in Astronomy Inc.\ under NASA contract NAS\,5-26555.  Some
 of the data presented herein were obtained at the W. M.\ Keck
 Observatory, which is operated as a scientific partnership among the
 California Institute of Technology, the University of California and
 the National Aeronautics and Space Administration.  The Observatory was
 made possible by the generous financial support of the W. M.\ Keck
 Foundation.  We are grateful to Mark Dickinson, Chuck Steidel, Lisa
 Storrie-Lombardi \& Ray Weymann for useful discussions, and Brenda Frye
 and Tom Broadhurst for providing details of their $z=4$ arcs in advance
 of publication. We have made use of the spectral evolutionary models of
 Gustavo Bruzual and St\'{e}phane Charlot. We thank Hans Hippelein and
 Klaus Meisenheimber at the Max-Planck-Institut f\"{u}r Astronomie for
 organizing an enjoyable and informative meeting at Ringberg.

\end{document}